\documentclass[usenatbib]{mn2e}
\usepackage{graphicx}
\usepackage{amssymb}
\newcommand{\Fs}{\,^*\! F}
\newcommand{\bA}{\bmath{A}}

\newcommand{\bJ}{\bmath{J}}
\newcommand{\bj}{\bmath{j}}

\newcommand{\bB}{\bmath{B}}
\newcommand{\bE}{\bmath{E}}
\newcommand{\bH}{\bmath{H}}
\newcommand{\bD}{\bmath{D}}

\newcommand{\bS}{\bmath{S}}

\newcommand{\text}[1]{\quad\mbox{#1}\quad}
\newcommand{\spr}[2]{\bmath{#1} \!\cdot\! \bmath{#2}}
\newcommand{\vpr}[2]{\bmath{#1} \!\times\! \bmath{#2}}
\newcommand{\vdiv}[1]{\spr{\nabla}{#1}}
\newcommand{\vcurl}[1]{\vpr{\nabla}{#1}}

\newcommand{\Pd}[1]{\partial_{#1}}
\newcommand{\Od}[1]{\frac{d}{d #1}}

\begin{document}

\title[3+1 Magnetodynamics]{3+1 Magnetodynamics}

\author[Komissarov S.S.]{S. S. Komissarov\thanks{E-mail: serguei@maths.leeds.ac.uk (SSK)}\\
$^{1}$Department of Applied Mathematics, The University of Leeds, Leeds, LS2 9GT, UK}

\maketitle

\begin{abstract}
The Magnetodynamics, or Force-Free Degenerate Electrodynamics, is recognised as 
a very useful approximation in studies of magnetospheres of relativistic stars. 
In this paper we discuss various forms of the Magnetodynamic 
equations which can be used to study magnetospheres of black holes. In 
particular, we focus on the 3+1 equations which allow for curved and dynamic 
spacetime. 
\end{abstract}


\section{Introduction}
\label{sec:intr}

In magnetospheres of pulsars and black holes the electromagnetic field is so 
strong that inertia and pressure of plasma can be ignored. As the result, 
the Lorentz force almost vanishes and the transport of energy and momentum is 
almost entirely electromagnetic \citep{GJ69,BZ77}.  This justifies the name 
``force-free'' to describe the electrodynamics of pulsars and black holes. 
However, the electrodynamics of the magnetospheres is rather different 
from the electrodynamics in vacuum which, obviously, is also force-free. 
Indeed, the magnetospheric plasma is plentiful enough to support 
strong electric currents and screen the component of electric field 
parallel to the magnetic field.  Electromagnetic field satisfying this condition 
is called ``degenerate'' and for this reason \citet{MT82} called the electrodynamics of 
pulsar and black hole magnetospheres ``force-free degenerate electrodynamics'' (FFDE). 

For a long time theorists were preoccupied with steady-state solutions of FFDE. 
Even the basic properties of FFDE as a system of time-dependent equation were not 
studied systematically. The first step in this direction wan made only quite recently, 
when \citet{U97} developed a theory of FFDE in which the electromagnetic field is 
described in terms of two scalar functions, called ``Euler potentials''.  However, this 
formulation has not been very popular. In particular, it is not very convenient 
for numerical analysis because its basic equations, when written in components, 
involve mixed space and time second order derivatives. Another approach is to 
use the actual Maxwell equations supplemented with a particular prescription for 
the electric current. This was done by \citet{G99}, who used the force-free condition to 
derive the Ohm law. \citet{K02} showed that FFDE can be considered 
as Relativistic MHD (RMHD) in the limit of vanishing inertia of plasma particles. This 
allowed to rewrite FFDE as a system of conservation laws similar to RMHD, including 
the energy-momentum conservation law. \citet{KBL07} argued that the dynamics of 
electromagnetic field in FFDE can be interpreted as a motion of magnetic 
mass-energy under the action of Maxwell stresses and proposed another name, 
``Magnetodynamics'' (MD), for FFDE. We will be using this name in the rest of the paper. 

The formulation by \citet{K02} is in a covariant form and can be used to study 
the magnetospheres of black holes \citep{K01,M06}. However, the wealth of 
experience accumulated in solving Maxwell equations has ensured that the formulation    
by \citet{G99} was often found preferable \citep{S06,KC09}. This prompted 
recent efforts to generalise Gruzinov's formulation so that it could also be used to 
study the magnetospheres of black holes. The starting point was the work by
\citet{TM82}, who first obtained general 3+1 equations of Electrodynamics 
(Eqs.(3.4) of this paper) and then a simplified version (Eqs.(5.8) of the same 
paper) which was adapted to the case of stationary black holes. 
Moreover, they restricted their attention to the Boyer-Lindquist foliation of spacetime. 
This simplified version has become most known to astrophysicists, via the follow-up 
paper by \citet{MT82} and \citet{TPM}, and widely used.     
\citet{K04} developed a different formulation, which has its roots in the works 
of \citet{Tamm} and \citet{Pleb} (see also \citet{LL71}). In this formulations 
the 3+1 equations of Electrodynamics also have a very simple and familiar form. 
In fact, they look exactly the same as the Maxwell equations in matter. 
The only assumption on the spacetime metric made in this formulation is 
that the determinant of the metric tensor of space does not depend on time. 
\citet{PLY10} presented, without derivation, the 3+1 equations 
which are free even from this constraint. They seem to have used the 
approach by \citet{K04} but reverted to the original representation of \citet{TM82}, 
where only the electric $\bE$ and magnetic field $\bB$ are present. Their equations 
also include extra scalar fields, which have been introduced for purely computational 
reasons.   
The force-free Ohm law of General Relativistic Magnetodynamics (GRMD) was first 
derived in the space-time  form by \citet{M06} and then in the 3+1 form by 
\citet{PBLR11}. \citet{L11} independently derived equations of GRMD using the 
simplified version of 3+1 Electrodynamics by \citet{TM82}. 
Thus, his equations have inherited the limitations of those by \citet{TM82}.    

In this paper, we revert back to the 3+1 formulation of \citet{K04}, modify
it in order to allow non-stationary metric,  
and derive the corresponding form of the force-free Ohm law.  We also 
present various relevant derivations and explore the connections 
between the different forms of the 3+1 equations.

\section{3+1 Electrodynamics} 
\label{sec-e}

Following \citet{TM82} we adopt the foliation
approach to the 3+1 splitting of spacetime in which    
the time coordinate $t$ parametrises a suitable filling of spacetime  
with space-like hypersurfaces described by the 3-dimensional metric tensor 
$\gamma_{ij}$. These hypersurfaces may be regarded as the 
``absolute space'' at different instances of time $t$. Below we 
describe a number of useful results for further references.     
If $\{x^i\}$ are the spatial coordinates of the absolute space then   
\begin{equation} 
  ds^2 = (\beta^2-\alpha^2) dt^2 + 2 \beta_i dx^i dt + 
         \gamma_{ij}dx^i dx^j \, ,
\label{metric} 
\end{equation}

\noindent
where $\alpha$ is called the ``lapse function'' and  
$\bbeta$ is the ``shift vector''. 
The 4-velocity of the local fiducial observer, 'FIDO',
which can be described as being at rest in the absolute space, is   

\begin{equation} 
   n_\mu = (-\alpha,0,0,0) \, .  
\label{ncov} 
\end{equation}

\noindent
The spatial components of the projection tensor, which is used to 
construct pure spatial tensors, 
\begin{equation} 
   \gamma_{\alpha\beta} = g_{\alpha\beta} + n_\alpha n_\beta \, ,
\end{equation}
coincide with the components of the spatial metric $\gamma_{ij}$.  
Other useful results are 

\begin{equation} 
   n^\mu = {1 \over \alpha}(1,-\beta^i) \, ,
\label{ncon} 
\end{equation}

\begin{equation} 
   g^{t \mu} = -{1 \over \alpha} n^\mu \, ,  
\label{gcov} 
\end{equation}

\begin{equation} 
   g = -\alpha^2 \gamma\, ,    
\label{dets} 
\end{equation}

\noindent
where 
$$
 \beta^i=\gamma^{ij}\beta_j  , \quad  
 g = \det{g_{\mu\nu}}, \quad 
 \gamma = \det{\gamma_{ij}} \, .  
$$
$\beta^i$ are the components of the velocity of the spatial grid relative  
to the local FIDO as measured using the coordinate time $t$ and the
spatial basis $\{ \Pd{i}\}$ \citep{MT82}.

The covariant Maxwell equations are  
(e.g. Jackson(1975)):

\begin{equation} 
  \nabla_\beta  \Fs^{\alpha \beta} = 0 \, , 
\label{Maxw1}
\end{equation}
and 

\begin{equation} 
  \nabla_\beta  F^{\alpha \beta} =  I^\alpha \, , 
\label{Maxw2}
\end{equation}

\noindent
where $F^{\alpha\beta}$ is the Maxwell tensor of the electromagnetic 
field, $\Fs^{\alpha\beta}$ is the Faraday tensor and  
$I^\alpha$ is the 4-vector of the electric current. 
The most direct way of 3+1 splitting of the covariant Maxwell equations 
is to write them down in components and then to introduce such spatial 
vectors that these equations have a particularly simple and familiar
form. For example, when Eq.(\ref{Maxw1}) is written in components it splits 
into two parts: 
\begin{itemize} 
\item{The time part:} 

\begin{equation} 
   \frac{1}{\sqrt{\gamma}} \Pd{i} 
   \left( \alpha\sqrt{\gamma} \Fs^{t i} \right) =0 \, ,       
\label{e1}
\end{equation}

\item{The spatial part:} 

\begin{equation}
   \frac{1}{\sqrt{\gamma}} \Pd{t} 
   \left( \alpha\sqrt{\gamma} \Fs^{j t} \right) +       
   \frac{1}{\sqrt{\gamma}} \Pd{i} 
   \left( \alpha\sqrt{\gamma} \Fs^{j i} \right) =0 \, .    
\label{e2}
\end{equation}

\end{itemize} 

\noindent
If we now introduce the spatial vectors $\bB$ and $\bE$ via 

\begin{equation} 
     B^i=\alpha \Fs^{it} \, ,
\label{B1}
\end{equation}
and 

\begin{equation}
   E_i =\frac{\alpha}{2} e_{ijk} \Fs^{jk} \, ,
\label{E1}
\end{equation}

\noindent
where 

\begin{equation} 
e_{ijk} = \sqrt{\gamma} \epsilon_{ijk}\, , \qquad 
e^{ijk} = \frac{1}{\sqrt{\gamma}} \epsilon^{ijk}, 
\end{equation}
is the Levi-Civita tensor of the absolute space and 
$\epsilon_{ijk}=\epsilon^{ijk}$ is the 3 dimensional Levi-Civita symbol, 
then equations (\ref{e1},\ref{e2}) read 

\begin{equation} 
   \vdiv{B}=0 \, , 
\label{divB} 
\end{equation}

\begin{equation} 
  \frac{1}{\sqrt{\gamma}} \Pd{t}(\sqrt{\gamma} \bB) + \vcurl{E} = 0 \, ,   
\label{Faraday}
\end{equation}

\noindent
where $\nabla$ is the covariant derivative of the absolute space.  
Similarly, equation (\ref{Maxw2}) splits into 

\begin{equation} 
\label{divD} 
   \vdiv{D}=\rho \, , 
\end{equation}

\begin{equation} 
   \frac{1}{\sqrt{\gamma}} \Pd{t}(\sqrt{\gamma} \bD) - \vcurl{H} = -\bJ \, ,
\label{Ampere}
\end{equation}

\noindent 
where 

\begin{equation} 
     D^i=\alpha F^{ti} \, ,
\label{D1}
\end{equation}

\begin{equation}
     H_i =\frac{\alpha}{2} e_{ijk} F^{jk} \, ,
\label{H1}
\end{equation}
and 

\begin{equation}
    \rho=\alpha I^t, \quad J^k=\alpha I^k \, . 
\label{rhoJ}
\end{equation}

Similar to any highly ionised plasma, the pair plasma of black hole 
magnetospheres has essentially zero electric and magnetic 
susceptibilities. In such a case, the Faraday tensor is simply 
dual to the Maxwell tensor   

\begin{equation} 
\Fs^{\alpha  \beta} = \frac{1}{2} e^{\alpha \beta \mu \nu} F_{\mu \nu} \, ,  
\label{dual_F}
\end{equation}

\begin{equation} 
F^{\alpha  \beta} = -\frac{1}{2} e^{\alpha \beta \mu \nu} \Fs_{\mu \nu} \, .   
\label{F}
\end{equation}
Here 

\begin{equation} 
e_{\alpha \beta \mu \nu} = \sqrt{-g}\,\epsilon_{\alpha \beta \mu \nu} \, ,  
\qquad
e^{\alpha \beta \mu \nu} = -\frac{1}{\sqrt{-g}}\,\epsilon^{\alpha \beta \mu \nu}  
\label{LCt}
\end{equation}

\noindent
is the Levi-Civita alternating tensor of spacetime 
and  $\epsilon_{\alpha \beta \mu \nu}=\epsilon^{\alpha \beta \mu \nu}$ 
is the four-dimensional Levi-Civita 
symbol. This allows us to obtain the following alternative expressions for 
$\bB,\bE,\bD$, and $\bH$: 
\begin{equation} 
     B^i= \frac{1}{2}e^{ijk} F_{jk} \, ,
\end{equation}

\begin{equation}
   E_i = F_{it} \, ,
\end{equation}

\begin{equation} 
     D^i = \frac{1}{2} e^{ijk} \Fs_{jk} \, ,
\end{equation}

\begin{equation}
     H_i = \Fs_{ti} \, . 
\end{equation}

\noindent
Moreover, from the above definitions one immediately finds the 
following vacuum constitutive equations: 

\begin{equation} 
     \bE = \alpha \bD + \vpr{\bbeta}{B} \, , 
\label{E3}
\end{equation}

\begin{equation} 
     \bH = \alpha \bB - \vpr{\bbeta}{D} \, . 
\label{H3}
\end{equation}
In flat spacetime with Lorentzian (pseudo-Cartesian) coordinates 
one has $\alpha=1$, $\beta=0$ and, hence, $\bB=\bH$ and $\bE=\bD$.   

Each of the introduced spacial vectors can be represented by a 
spacetime vector whose spacial part is the spacial vector in question and whose 
time part vanishes. As one can easily verify, these spacetime vectors are given by the 
following covariant expressions: 

\begin{equation} 
     B^\mu = -\Fs^{\mu\nu} n_\nu \, ,
\label{B2}
\end{equation}

\begin{equation} 
     E^\mu = \frac{1}{2}\gamma^{\mu\nu} e_{\nu\alpha\beta\gamma} 
          k^{\alpha} \Fs^{\beta\gamma} \, ,  
\label{E2}
\end{equation}

\begin{equation} 
     D^\mu = F^{\mu\nu} n_\nu \, ,
\label{D2}
\end{equation}

\begin{equation} 
     H^\mu = -\frac{1}{2}\gamma^{\mu\nu} e_{\nu\alpha\beta\gamma} 
             k^{\alpha} F^{\beta\gamma} \, ,  
\label{H2}
\end{equation}

\begin{equation} 
     J^\mu = 2I^{[\nu}k^{\mu]}n_\nu \, , 
\label{J2}
\end{equation}
where $ k^\alpha = \Pd{t}$. From these one can see that $\bB$ and $\bD$ 
are the magnetic and electric fields as measured by FIDOs, whereas 
$\bH$ and $\bE$ are auxiliary vector fields.  

It is also easy to verify that 

\begin{equation}
    \rho=-I^\nu n_\nu,
\label{rho2}
\end{equation}
and thus $\rho$ is the electric charge density as measured by FIDOs. 
However, $\bJ$, is not the electric current density as measured by 
FIDO, which we will denote as $\bj$. Geometrically, $\bj$ is the component 
of $I^\nu$ normal to $n^\nu$. Using the projection tensor 
$\gamma_{\nu\mu}=g_{\nu\mu}+n_\nu n_\mu$, we find    

\begin{equation} 
     \bJ=\alpha\bj-\rho\bbeta \, .
\label{e6}
\end{equation}
The second term in this equation accounts for the motion of spatial 
grid relative to FIDO, or in other words for the fact that the coordinate 
time direction, the basis vector $\Pd{t}$, is generally not parallel to $n^\nu$.

When $\Pd{t}\gamma=0$ these 3+1 equations have exactly 
the same form as the classical Maxwell equations for the electromagnetic field in 
matter 

\begin{equation}
   \vdiv{B}=0\, ,
\label{divB-1}
\end{equation}

\begin{equation}
   \Pd{t}\bB + \vcurl{E} = 0\, ,
\label{Faraday-1}
\end{equation}

\begin{equation}
\label{divD-1}
   \vdiv{D}=\rho\, ,
\end{equation}

\begin{equation}
   -\Pd{t}\bD + \vcurl{H} = \bJ\, . 
\label{Ampere-1}
\end{equation}
This similarity explains why we prefer to denote the electric field measured 
by FIDO as $\bD$, whereas in most papers by other researches it is denoted as 
$\bE$.

Applying the divergence operator to Eq.\ref{Ampere} one finds the electric 
charge conservation law

\begin{equation}
   \frac{1}{\sqrt{\gamma}} \Pd{t}(\sqrt{\gamma} \rho) + \vdiv{J} = 0 .
\label{rho-cons-diff}
\end{equation}
Although this is slightly different from the usual differential form of 
this law, its integral form is exactly the same 

\begin{equation}
   \Od{t} \int\limits_V \rho dV + \int\limits_S \spr{J}{dS} = 0, 
\label{rho-cons-int}
\end{equation}
where $dV$ is the metric volume and $d\bS$ is the metric surface elements.

The limit of Magnetodynamics is defined by vanishing of the Lorentz force. 
In the covariant form this condition reads as 

\begin{equation}
   F_{\mu\nu} I^{\mu}=0\, .  
\label{FFC}
\end{equation} 
In our 3+1 formulation this equation splits into  

\begin{equation}
   \spr{E}{J}=0 
\label{a2} 
\end{equation} 
and 

\begin{equation}
  \rho \bE + \vpr{J}{B} =0\, .
\label{a3} 
\end{equation} 
From the last equation it follows that

\begin{equation}
   \spr{\bE}{\bB}=0\, .
\label{EB}
\end{equation}
When combined with the constitutive equation (\ref{E3}), the last 
equation also implies 
\begin{equation}
   \spr{\bD}{\bB}=0\, .
\label{DB}
\end{equation}
As first noticed by \citet{G99}, the force-free condition allows one to express 
the electric current in terms of the electromagnetic field and its spacial 
derivatives, thus providing us with a particular form of Ohm's law. 
Here we repeat Gruzinov's derivation taking into account the effects of General Relativity.  
The component of electric current normal to the magnetic field can be found 
directly from Eq.\ref{a3} via cross-multiplying its sides by $\bB$. 
This yields 

\begin{equation}
    \bJ_\perp = \rho \frac{\vpr{E}{B}}{B^2}  \, .
\label{J-perp}
\end{equation}
In order to find the parallel component we first notice that Eq.\ref{DB} 
implies
 
\begin{equation}
  \Pd{t}(\sqrt{\gamma} \spr{\bD}{\bB})=0\, .
\end{equation}
When combined with Eqs.(\ref{Faraday},\ref{Ampere},\ref{DB}) this yields 

\begin{equation}
   (\vcurl{H}-\bJ)\cdot\bB-(\vcurl{E})\cdot\bD=0\, , 
\end{equation} 
which does not involve the time derivative of $\gamma$. From the last result 
we find that  

\begin{equation}
    \bJ_\parallel = \frac{\bB\cdot(\vcurl{H})-\bD\cdot(\vcurl{E})}{B^2} \bB \, .
\label{J-parall}
\end{equation}

Collecting all these results, we can write the most general 3+1 system of 
GRMD as 

\begin{equation}
   \frac{1}{\sqrt{\gamma}} \Pd{t}(\sqrt{\gamma} \bB) + \vcurl{E} = 0 \, ,
\label{Faraday-f}
\end{equation}

\begin{equation}
   \frac{1}{\sqrt{\gamma}} \Pd{t}(\sqrt{\gamma} \bD) - \vcurl{H} = -\bJ \, ,
\label{Ampere-f}
\end{equation}

\begin{equation}
   \vdiv{B}=0\, ,
\label{divB-f}
\end{equation}

\noindent
where

\begin{equation}
     \bE = \alpha \bD + \vpr{\bbeta}{B}\, ,
\label{E-f}
\end{equation}

\begin{equation}
     \bH = \alpha \bB - \vpr{\bbeta}{D}\, ,
\label{H-f}
\end{equation}

\begin{equation}
    \bJ = \rho\frac{\vpr{E}{B}}{B^2} + 
\frac{\bB\cdot(\vcurl{H})-\bD\cdot(\vcurl{E})}{B^2} \bB \, ,
\label{J-f}
\end{equation}
and

\begin{equation}
\label{divD-f}
   \rho=\vdiv{D}\, .
\end{equation}

It is easy to see that in flat spacetime with Lorentzian coordinates, where 
$\alpha=1$, $\beta=0$, and $\Pd{t}\gamma=0$, this system is reduced to that of 
\citet{G99}. Under the conditions $\Pd{t}\gamma=0$ and $\vdiv{\beta}=0$ it is 
reduced to that of \citet{L11}. 

Magnetodynamics can be considered as Relativistic Magnetohydrodynamics  in the 
limit of vanishing particle inertia \citep{K02}. The explicit condition of  
magnetohydrodynamic approximation is vanishing of the electric field in the 
fluid frame. This implies that in any other frame the component of electric field 
parallel to the magnetic one always vanishes and the magnetic field is stronger 
than the electric one. These conditions can be written in the covariant form as 

\begin{equation}
  \Fs_{\mu\nu}F^{\mu\nu}=0 \text{and} 
  F_{\mu\nu}F^{\mu\nu}>0 \, .
\end{equation} 
In our 3+1 notation these yield 

\begin{equation}
   \spr{B}{D}=0 \text{and} B^2-D^2 >0\, .
\label{a11} 
\end{equation} 
In computer simulation, one has to make sure that the initial 
solution satisfies both these conditions. The first constraint is preserved 
exactly by the differential equations of MD. However, the second 
constraint can be violated \citep{K02}. Slow shocks of RMHD can 
transform plasma from magnetically-dominated to particle-dominated state \citep{Lb05}. 
However, slow waves are not allowed in the MD approximation \citep{K02}. 
This limitation can be behind many violations of the second 
condition (\ref{a11}) in MD.

Substituting the expressions for $\bE$ and $\bH$ from the constitutive equations 
into Eqs.(\ref{Faraday},\ref{Ampere}) and expanding the double cross-product terms
one finds 
\begin{equation}
   \Pd{t}\bB -{\cal L}_\bbeta \bB + \vcurl{\alpha \bD} = 
   \eta\bB 
\label{e4a}
\end{equation}
and
\begin{equation}
   \Pd{t}\bD -{\cal L}_\bbeta \bD - \vcurl{\alpha \bB} = 
   \eta\bD -\alpha\bj \, ,
\label{e5a}
\end{equation}   
where $\eta=\vdiv{\beta}-\Pd{t}(\ln\sqrt{\gamma})$
and ${\cal L}_\bbeta $ is the Lie derivative along the shift vector 
(e.g. ${\cal L}_\bbeta \bB = (\spr{\beta}{\nabla})\bB - (\spr{B}{\nabla})\bbeta $ ). 
This is another useful form of the most general Faraday and Amp\'ere equations of 
3+1 GR Electrodynamics \citep{PLY10}\footnote{In \citet{PLY10}, 
as well as in \citet{TM82} and many other papers, 
the variable $\bD$ is denoted as $\bE$, following its interpretation as the electric 
field measured by the local FIDO of the spacetime foliation.}. 
One can show that 

\begin{equation} 
\eta = \alpha Tr(K), 
\end{equation}
where $Tr(K)=\gamma^{ik}K_{ik}$ is the trace of the external curvature tensor 
of the absolute space 

\begin{equation} 
  K_{ik} = \frac{1}{2\alpha} \left( \beta_{i;k}+\beta_{k;i}-\Pd{t} \gamma_{ik} \right)
\end{equation}
\citep{M73}.
When both  $\Pd{t}\gamma=0$ and  $\vdiv{\beta}=0$  these equations reduce to  

\begin{equation} 
   \Pd{t}\bB -{\cal L}_\bbeta \bB + \vcurl{\alpha \bD} = 0 
\label{e4}
\end{equation}
and
\begin{equation} 
   -\Pd{t}\bD +{\cal L}_\bbeta \bD + \vcurl{\alpha \bB} = \alpha\bj \, . 
\label{e5}
\end{equation}
These are the 3+1 equations of Black Hole Electrodynamics by
\citet{MT82}. We note here that although the condition $\vdiv{\beta}=0$ 
is satisfied by the Boyer-Lindquist metric of Kerr black holes it is not 
satisfied by the Kerr-Schild  metric, which is also widely used in black hole 
studies. 

In terms of the physical quantities measured by FIDOs, vanishing of the Lorentz 
force has the familiar form\footnote{This equation can also be obtained via 
substituting expressions (\ref{E3}) and (\ref{e6}) into Eq.(\ref{a3}).}
\begin{equation}
  \rho \bD + \vpr{j}{B} =0\, .
\label{n1}
\end{equation} 
The force-free electric current $\bj$ can now be obtained in exactly the same 
fashion we did earlier for $\bJ$. The normal component of $\bj$ is obviously  

\begin{equation}
    \bj_\perp = \rho \frac{\vpr{D}{B}}{B^2} \, .
\label{j-perp}
\end{equation}
In order to find the parallel component we apply the operator 
$\Pd{t} -{\cal L}_\bbeta$ to $\spr{B}{D}=0$. This yields 

\begin{equation}
    \bj_\parallel = \frac{\bB\cdot(\vcurl{\alpha B})-
                     \bD\cdot(\alpha\vcurl{D})}{\alpha B^2} \bB \, .
\label{j-parall}
\end{equation}
Given the identity 
$\bA\cdot(\vcurl{\alpha A})\equiv\alpha\bA\cdot(\vcurl{A})$  
the final expression for the force free current does not 
actually involve either the shift vector or the lapse function
and has exactly the same form as in Special Relativity, 

\begin{equation}
    \bj =  \rho \frac{\vpr{D}{B}}{B^2} + 
     \frac{\bB\cdot(\vcurl{B})-\bD\cdot(\vcurl{D})}{B^2} \bB \, 
\label{j}
\end{equation}
\citep{PBLR11}.

\section{The 4-vector of force-free current}

Finally, we briefly discuss the space-time formulation of Magnetodynamics. 
If one prefers to deal with the 4-tensor Maxwell-Amp\'ere equation (\ref{Maxw2})
instead of the energy-momentum equation (as in \citet{K02}) then  
the key issue is the expression for the 4-vector of force-free current. 
This expression was found by \citet{M06}. However, it can be simplified 
a little bit further. Here we explain this and give a slightly different 
derivation.

From the definitions (\ref{B2},\ref{D2}) it follows that 

\begin{equation}
    F^{\alpha\beta}=
    n^\alpha D^\beta -D^\alpha n^\beta - e^{\alpha\beta\nu\xi}B_\nu n_\xi \, ,
\label{FDB}
\end{equation}
and 

\begin{equation}
    \Fs^{\alpha\beta}=
    -n^\alpha B^\beta + B^\alpha n^\beta - e^{\alpha\beta\nu\xi}D_\nu n_\xi \, .
\label{FsDB}
\end{equation}
Then the force-free condition (\ref{FFC}) reads 

\begin{equation}
  \rho D^\beta + e^{\xi\beta\alpha\nu} n_\xi I_\alpha B_\nu  =0. 
\label{FFC1}
\end{equation}
From this we find that 

\begin{equation}
  D^\beta B_\beta =0 
\label{DB0}
\end{equation}
and 

\begin{equation}
  I^\mu = \frac{\rho}{B^2} e^{\gamma\mu\beta\delta} n_\gamma D_\beta B_\delta +
         \frac{(I^\nu B_\nu)}{B^2} B^\mu +\rho n^\mu \, .
\label{I}
\end{equation}
One can see that the spacial part of $I^\mu$, which we will denote as 
${\cal J}^\mu={\cal J}^\mu_\parallel+ {\cal J}^\mu_\perp $, 
has the following components parallel and perpendicular to $B^\mu$ 

\begin{equation}
  {\cal J}^\mu_\parallel = \frac{(I^\nu B_\nu)}{B^2} B^\mu \, ,
\label{Jpa}
\end{equation}

\begin{equation}
 {\cal J}^\mu_\perp = 
 \frac{\rho}{B^2} e^{\gamma\mu\beta\delta} n_\gamma D_\beta B_\delta \, .
\label{Jpe}
\end{equation}

The coefficient $I^\nu B_\nu$ in Eq.\ref{Jpa} can be expressed in terms of 
the electric and magnetic fields and their derivatives, making this equation 
an explicit expression for ${\cal J}^\mu_\parallel$. Following \citet{M06} we
first contract the Maxwell-Amp\'ere law (\ref{Maxw2}) with $B^\mu$ to find 
that 

\begin{equation}
   I^\alpha B_\alpha = 
                      -B^\alpha D_{\alpha,\beta} n^\beta  
                      -e^{\alpha\beta\nu\xi} B_\alpha B_{\nu,\beta} n_\xi \, , 
\end{equation} 
where the comma indicates partial derivative.
Then we contract the Maxwell-Faraday equation (\ref{Maxw1}) with $D_\nu$ to 
find that 
\begin{equation}
 B^\alpha D_{\alpha,\beta} n^\beta= 
                      -e^{\alpha\beta\nu\xi} D_\alpha D_{\nu,\beta} n_\xi \,.
\end{equation}
Thus, 

\begin{eqnarray}
   I^\alpha B_\alpha =
        e^{\xi\alpha\beta\nu} n_\xi 
          (B_\alpha B_{\nu,\beta} - D_\alpha D_{\nu,\beta} ) =
\nonumber
\end{eqnarray} 
\begin{equation}
     \qquad\qquad = 
        e^{\xi\alpha\beta\nu} n_\xi 
          (B_\alpha B_{\nu;\beta} - D_\alpha D_{\nu;\beta} ) \, ,
\label{IB}
\end{equation} 
where the semi-colon stands for covariant differentiation. The corresponding 
expression in \citet{M06} is a little bit different because it includes the 
term $ B^\alpha D^\beta(n_{\beta;\alpha}+n_{\alpha;\beta})$, which equals to zero.  
Collecting all the results, we obtain 

\begin{eqnarray}
 {\cal J}^\mu = 
 \frac{\rho}{B^2} e^{\gamma\mu\beta\delta} n_\gamma D_\beta B_\delta\, +
\nonumber
\end{eqnarray} 
\begin{equation}
        \qquad\qquad B^{\mu} \frac{e^{\xi\alpha\beta\nu} n_\xi 
          (B_\alpha B_{\nu;\beta} - D_\alpha D_{\nu;\beta} )}{B^2} \, ,
\label{J}
\end{equation} 
\begin{equation}
        I^\mu = \rho n^\mu+ {\cal J}^\mu \, .
\end{equation}
It is easy to verify that in the 3+1 notation Eq.\ref{J} is identical to Eq.\ref{j},
which does not include neither the lapse function nor the shift vector, nor 
the time derivatives of $\bB$ and $\bD$.

\section{Acknowledgments}
It is a pleasure to thank Maxim Lyutikov and Jonathan McKinney for 
stimulating discussions.


\end{document}